\documentclass[conference]{IEEEtran}

\usepackage[pdftex]{graphicx}
\usepackage[autostyle]{csquotes}
\usepackage{enumitem}
\usepackage{multicol}
\usepackage{tabu}
\usepackage{caption}
\usepackage{listings}
\usepackage{array, booktabs, ragged2e}
\begin{document}

\title{Portable, high-performance containers for HPC}

\author{\IEEEauthorblockN{Lucas Benedicic\IEEEauthorrefmark{1},
Felipe A. Cruz,
Alberto Madonna,
Kean Mariotti}
\IEEEauthorblockA{Systems Integration Group\\
CSCS, Swiss National Supercomputing Centre\\
Lugano, Switzerland\\
Email:
\IEEEauthorrefmark{1}lucas.benedicic@cscs.ch
}
}

\maketitle

\begin{abstract}

Building and deploying software on high-end computing systems is a challenging task. High performance applications have to reliably run across multiple platforms and environments, and make use of site-specific resources while resolving complicated software-stack dependencies. Containers are a type of lightweight virtualization technology that attempt to solve this problem by packaging applications and their environments into standard units of software that are: portable, easy to build and deploy, have a small footprint, and low runtime overhead.
In this work we present an extension to the container runtime of Shifter that
provides containerized applications with a mechanism to access GPU accelerators and specialized networking from the host system, effectively enabling performance portability of containers across HPC resources.
The presented extension makes possible to rapidly deploy high-performance software on supercomputers from containerized applications that have been developed, built, and tested in non-HPC commodity hardware, e.g. the laptop or workstation of a researcher.

\end{abstract}

\IEEEpeerreviewmaketitle

\section{Introduction}

Software portability is a long-time concern for software development. One of the first initiatives that tried to tackle the complexity and cost of porting software to new platforms took place in the late 1950s with the creation of the COBOL programming language \cite{Ensmenger2010}.
More than 60 years have passed since then and we are still struggling with the same fundamental challenge.

Portability is also a great concern in the area of High Performance Computing (HPC) as most often HPC applications have to run across many platforms and environments while achieving high levels of computational performance.

Throughout this work we will use the definition of software portability provided in \cite{mooney1990strategies}:

\blockquote{A software unit is portable (exhibits portability) across a class of environments to the degree that the cost to transport and adapt it to a new environment is less than the cost of redevelopment.}

Since the initial work in COBOL many new innovations have inched us closer towards addressing portability of software over the years. Consider the progress on the following areas:
programming languages (e.g. COBOL, Fortran, C, C++, Python),
portable libraries (e.g. Boost C++, PETSc),
universal operating systems (e.g. Linux from embedded systems to supercomputers),
computer platform standards (e.g. IBM-PC, x86-based architectures),
software patterns (e.g. portability through reusable software components).
Although these efforts have become important tools at the disposal of software developers, portability is a problem that is compounded by other pressing requirements, for instance: performance, resource limitations, machine-dependent features, OS-dependent requirements, availability of libraries and tools.
Hence, it is often the case for developers to address portability as a tradeoff with other requirements.
Consider for example the approach used by Hadoop, an open-source framework for distributed processing and storage of large datasets, where portability is achieved by abstracting away the details of the platform's software and hardware.
As discussed in \cite{shafer2010hadoop}, the software abstraction used by Hadoop clearly incurs in a tradeoff between performance and portability.

In the context of HPC, application portability and high-performance need to be achieved together. This is especially challenging considering that supercomputers architecture, configuration, and software environments greatly differ between systems and vendors.
The diversity of environments alone make necessary for software to be able to adjust to each system.
All things considered, the task of porting code across supercomputer sites is costly and time-consuming.
However, most often users and developers would like to spend their time working in new science and software features rather than porting applications between platforms and environments.
Therefore, simplifying and accelerating the workflow for porting applications provides great gains in the productivity of users and developers.

Virtualization, a technology that has seen rapid growth over the last decade, has shown potential to ease portability and deployment of applications particularly in Cloud environments.
Virtualization as a term is used for technologies that improve portability and maintainability of software by separating part or all of the operating system (OS) and software environment from the underlying hardware where the application will be run.
Computer virtualization allows then for virtualized software to run without modification on other systems, enabling a faster software-porting workflow.
Until recently, the adoption of virtualization on HPC settings has been limited due to concerns related to the performance of virtualized applications~\cite{walters2008comparison}. However, some of these issues have been addressed by container virtualization~\cite{morabito2015hypervisors}\cite{Benedicic_2016}.

Container virtualization is seen as a more flexible, lightweight, and easier to deploy alternative to hypervisor-based virtualization.
A hypervisor provides portability by virtualizing the hardware and requiring a full OS to run inside the virtual machine that will run an application. In contrast, container virtualization will run a container instance natively on top of the host system OS kernel.

In this work we discuss an approach to container-based virtualization that addresses the usage of specialized HPC resources such as GPU accelerators and network interconnect solutions.
We have implemented this by extending Shifter~\cite{Jacobsen_Canon_2016}, a project that provides a container runtime built specifically to address the needs of HPC.

This paper is organized as follows: Section~\ref{sec:background} provides a brief overview of the different technologies that enable this work. Section~\ref{sec:portableHPCcontainers} describes Shifter, its internal and user workflows. Our contributions are presented in Section~\ref{sec:Extending-Shifter}. In Section~\ref{sec:performanceEvaluation} we present the methodology and benchmarks used to validate portability and performance. Finally, Section~\ref{sec:conclusion} provides the conclusion.

\section{Background}\label{sec:background}

In this section, we provide the technological context that enables Shifter's performance portability. We briefly introduce the main concepts behind containers, interfaces, and standards.

\subsection{Containers}\label{subsec:background-containers}

Containers are a type of virtualization that takes place at the OS level. Containers communicate with the host OS through system calls to the kernel.
Therefore, the OS kernel is the interface layer between user-space containerized applications and the hardware resources of the host system that the application accesses during deployment.

Interfacing with the Linux kernel bestows a high degree of portability to containers, as it provides the same interfaces and behavior independently of the hardware platforms it is executed on.
This ensures that containerized applications can be seamlessly deployed across diverse hardware platforms.
However, the fact that container images are hardware agnostic means that hardware-accelerated features which require specialized drivers, libraries, and tools are not commonly part of a container image.
Therefore, providing access to these specialized resources is crucial for bringing container technology to HPC, allowing containers to be seamlessly deployed across diverse platforms.

The container image is a standardized unit of software that packages the application and all the dependencies needed for its correct execution.
Even though the container itself is stateless and does not store persistent data, it can still access persistent data by connecting to a data storage volume that is independent from the container's life cycle.
In an HPC environment, it is most often provided by a parallel file system like Lustre or BeeGFS.

Linux containers and containerization are general terms that refer to the implementation of OS-level virtualization.
Moreover, there exists multiple such implementations and may refer to one of the following technologies: Docker~\cite{docker}, LXC~\cite{lxc}, Rocket~\cite{rkt}, Shifter~\cite{Jacobsen_Canon_2016}, Singularity~\cite{kurtzer_2016_60736}, etc.
Nevertheless, throughout this work we exclusively discuss the use of Docker container images and the Shifter container runtime. Docker container images are the de-facto standard for defining the encoding of a container, i.e., the bundle of metadata, directory structure, and files that taken together compose the containerized application.
The container runtime is then responsible of managing the kernel features and system resources that a containerized applications can use, thus the runtime grants the container with access to the host resources (devices, drivers, libraries, and tools).

Containers, in addition to aiding the portability efforts, also bring other benefits to HPC users that are worth noting \cite{Jacobsen_Canon_2016}\cite{hale2016containers}\cite{julian2016containers}\cite{Felter_2015}\cite{boettiger2015introduction}. As a summary, consider the following:
\begin{itemize}

\item Encapsulation of complete software environments.

\item Packaging of applications with different software versions and dependencies, avoiding software with conflicting versions on the host.

\item Simplified collaboration, since containers as an unit of software can be easily shared.

\item Improved reproducibility, since containers are self-contained and can be used to replicate results consistently.

\item Improved portability of applications, containers run with the same configuration across different environments.

\item Improved development flexibility, containers allow development on non-HPC environments.

\item Rapid deployment and execution at scale.

\end{itemize}

\subsection{Interfaces and standards}

The use of standards and interfaces is at the core of the portability of containers.
On the one hand, standards are widely accepted specifications that define and describe the characteristics and behavior of a function, object, or system.
As such, the use of standards is pivotal for the success of efforts to achieve portable applications.
Standards are used to define interfaces that can be supported by different environments.
On the other hand, interfaces play a central role to portability by providing clear abstractions that enable software components to interact at different levels and environments.

There are two types of interfaces:
one is the application programming interface (API) through which two software components can interact at the source code level;
the other one is the application binary interface (ABI), which defines the interaction of software components at a machine-code level.
Moreover, the ABI guarantees binary compatibility between object modules, i.e., a software component can interact with any other ABI-compatible component without recompilation.  
It is then through the use of the API and the ABI that libraries implementing these interfaces can achieve interoperability between vendors and platforms.

Our extension of Shifter makes use of libraries that are compatible through their ABI to make site-optimized system features available while maintaining the portability that container images provide.
Our implementation solves the practical details to transparently enable container portability and native performance on a variety of target systems through a simple and straightforward usage of command-line and configuration options.

We will now discuss the most often used standards to access two types of specialized hardware commonly available on supercomputer sites: the Message Passing Interface (MPI) for fast network interconnects and Nvidia's CUDA runtime API for accessing GPU compute accelerators.

\subsubsection{MPI library}\label{subsubsec:background-interfaces-mpi}

The MPI \cite{MPI30} is the de-facto industry standard for developing distributed memory applications.
HPC centers around the globe extensively use implementations of MPI that can take advantage of the fast network infrastructure while providing communications that are low-latency, efficient, and scalable.

Version 1.0 of the MPI standard was introduced in 1994 and presented a set of low-levels routines that cover point-to-point and collective communications.
MPI is both an API and a specification for how the implemented routines must behave.
Moreover, MPI does not define the implementation details of the routines, allowing vendors to build specialized libraries that can efficiently use the hardware acceleration developed for a particular hardware.

From its inception, MPI provided portability at the API level.
However, in 2013 multiple MPI vendors announced an effort to achieve ABI compatibility between their implementations~\cite{MPIABI}.
In practice, it is the ABI compatibility that allows Shifter, under user request, to switch the MPI implementation from the one included in the container with the specialized version that is available on the host system.
This allows the container to access an optimized MPI library that can take advantage of hardware acceleration.

\subsubsection{CUDA}\label{subsec:backgroud-cuda}

CUDA stands for Compute Unified Device Architecture and is the API for NVIDIA GPU accelerators~\cite{CUDACPROGRAMMING}. It is the de-facto standard programming framework developed used to scale applications on large GPU systems efficiently.
The NVIDIA CUDA C compiler can produce forward-compatible GPU code versions (in PTX format) of an application, this ensures that the application remains compatible with future versions of the CUDA runtime.
It is this version compatibility that Shifter exploits to give containers access to the host system GPUs, libraries, and CUDA related binaries with the ones available on the host system at runtime.

\section{Shifter}\label{sec:portableHPCcontainers}

In the scientific computing community there has been a growing interest in running Docker containers on HPC infrastructure~\cite{deal2016hpc}.
 However, Docker was developed to answer the particular needs of web service applications, which feature substantially different workloads from the ones typically encountered in HPC.
Shifter is a software solution meant to bridge this gap, allowing containerized applications to integrate completely into an HPC infrastructure and to run with full hardware capabilities.

\begin{figure}[h]
	\centering
	\includegraphics[width=0.46\textwidth]{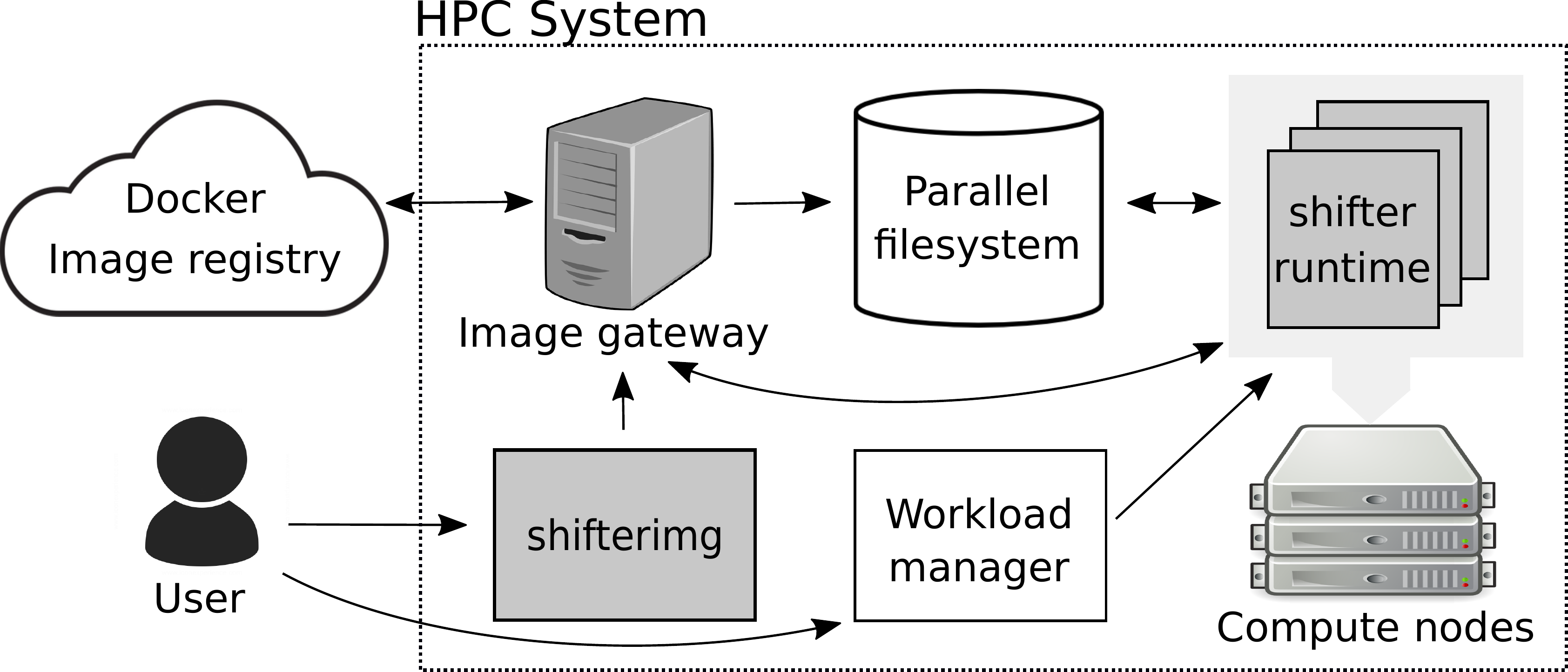}
	\caption{Architecture overview of Shifter. Highlighting the Image Gateway and the Shifter Runtime.}
	\label{fig:shifter-diagram}
\end{figure}

Fig.~\ref{fig:shifter-diagram} shows the general architecture of Shifter, highlighting the Image Gateway and Runtime components.

The Image Gateway manages container images, importing (or \emph{pulling}) from remote Docker registries and converting them to Shifter's custom format at the user's request. The Gateway also maintains the images in a location accessible system wide, and allows the user to query the currently available images.

Shifter is a lean runtime that has been tailored for running containers on high performance infrastructure while addressing the needs of supercomputing environments:

\begin{enumerate}
\item maintaining user privileges during execution;
\item providing a scalable solution for distributed workloads;
\item maintaining compatibility with older Linux kernels;
\item granting security by avoiding the use of a root daemon process to run containers;
\item workload manager integration (e.g. SLURM and ALPS);
\item integrating with parallel file systems.
\end{enumerate}

The rest of this Section describes the procedure Shifter implements to import a Docker image and execute a containerized application.
We then outline the user workflow enabled by Shifter, moving from a personal workstation to a container deployed on an HPC system.

\subsection{Shifter Overview}

The execution of a container on a host system through Shifter can be broken down into several stages.
The Image Gateway performs only the first phase of the workflow (pulling and reformatting images), while Shifter's Runtime is responsible for all the others.

\begin{description}[leftmargin=!]

\item[Pulling and reformatting of images] The Image Gateway is instructed by the user to pull a container image from a remote Docker registry. The Gateway queries the registry for the image matching an identifier supplied by the user, and starts the download in a temporary directory.

Once the Image Gateway is finished pulling, the Docker image is immediately expanded, flattened (i.e., all layers but the last one are discarded), converted to a squashfs format (a compressed real-only file system for Linux), and finally stored into the parallel file system.

Container images processed by the Image Gateway can be queried or listed, and are ready for execution by the Runtime.

\item[Preparation of software environment] It is carried out by the Runtime prior to the execution of a container application from an image available on the Image Gateway.
The objective of this setup stage is to prepare the software environment that will be present inside the container during its execution.

Shifter builds the container environment from two sources: the user-specified image and the parts of the host system Shifter has been configured to source.
In practice, the squashfs image is copied to the compute nodes and its file system bundle (that packs the application and all its dependencies) is appropriately loop mounted into a local directory, which will become the root directory of the container.
The file structure is augmented with site-specific resources from the host system (e.g., parallel filesystem directories, GPU devices, network devices) and site-specific tools and libraries.

\item[Chroot jail] This stage changes the root directory of the container environment to the directory structure that was prepared during the previous stage, effectively isolating the container filesystem from the one of the host system.

\item[Change to User and Group privileges] At this point Shifter has completed the steps that require additional system privileges, namely the setup of the container environment and the change of its root directory. Shifter now drops the extra privileges, preparing to run the container application under normal user and group privileges. This is done by setting the effective user id and group id with the \texttt{setegid()} and \texttt{seteuid()} system calls respectively. 

\item[Export of environment variables] The environment variables from the image are copied over to the container environment.
In order for the container to correctly support the host-specific resources, selected variables from the host system are also added to the environment, according to Shifter configuration files.
  
\item[Container application execution] Shifter now executes the application as the end user: in this way, the user's privileges and accounting information are maintained, as would happen with any other process running on the host system.

\item[Cleanup] Once the application exits, the resources used to create the container software environment are released.

\end{description}

\subsection{User workflow}

\begin{figure}[h]
	\centering
	\includegraphics[width=0.4\textwidth]{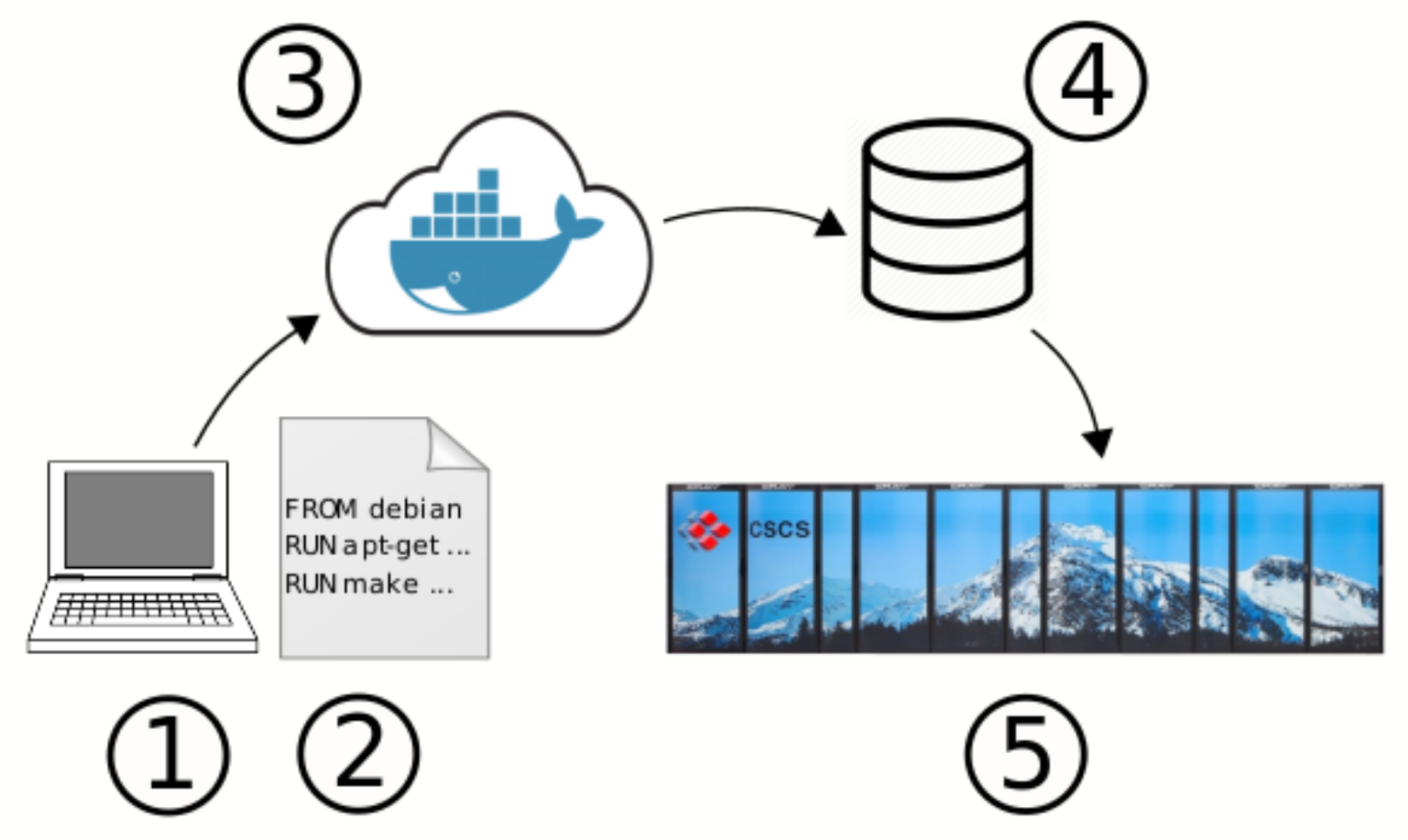}
	\caption{User workflow with Docker and Shifter. Brief summary of the workflow steps: 1) build container; 2) test container; 3) Push container to Docker registry; 4) Pull image into Shifter; 5) Run container with Shifer.}
	\label{fig:shifter-user-workflow}
\end{figure}

A typical Docker workflow starts with a container image built on a workstation, which is then uploaded to the cloud in order to be shared and deployed elsewhere.

Although the details of how Shifter builds the container software environment and runs a containerized application might seem involved, the end user workflow for running a container with Shifter closely resembles the one used by Docker, with the advantage that Shifter enables the deployment to happen at scale on a HPC system.
On such system, the user has to work with the command line utilities provided by Shifter, namely \texttt{shifterimg} and \texttt{shifter}:

\begin{itemize}
	\item \texttt{shifterimg} provides a way for the user to interact with the Image Gateway, in order to request image pulls from a Docker registry or to query available images.
	\item \texttt{shifter} is the application that constitutes the Runtime, being responsible of building the container software environment and launching a containerized application.
\end{itemize}

The end-user workflow of combining Docker and Shifter for running containers on HPC infrastructure is depicted in Fig.~\ref{fig:shifter-user-workflow} and can be summarized as follows:

\begin{enumerate}

\item Build a container image using Docker. This generally happens on a user's personal machine. The availability of popular Linux operating systems as base images offers familiar tools for managing dependencies and building software inside the image.

\item Test the image locally to verify the software stack and applications work correctly.

\item Push the container image to a remote registry, e.g., http://hub.docker.com

\item Pull the image into an HPC system with Shifter Image Gateway, using \texttt{shifterimg}.

\item Run the container using Shifter Runtime \texttt{shifter}.
Even though the applications were built on a workstation, with no information about the final target platform, they should run on HPC hardware.

\end{enumerate}

As it can be observed above, steps (4) and (5) involve Shifter. For illustrative purposes consider the following example: on a Cray XC50 System, Shifter is used to execute a Docker image already available from http://hub.docker.com. We arbitrarily choose the Docker image for Ubuntu 16.04 "Xenial Xerus".

To pull the image with Shifter we use the image identifier:

\begin{lstlisting}
shifterimg pull docker:ubuntu:xenial
\end{lstlisting}

Once the image has been pulled and processed by the Image Gateway, it can be used by the Runtime to create a container and execute a command.
In this case, we ask the container to print information relative to its OS, to effectively verify that it is working in the Ubuntu environment packaged in the image:

\begin{lstlisting}
shifter --image=ubuntu:xenial cat /etc/os-release
\end{lstlisting}

The output of which on the Cray XC50 system is:

\begin{lstlisting}
NAME="Ubuntu"
VERSION="16.04.2 LTS (Xenial Xerus)"
ID=ubuntu
ID_LIKE=debian
PRETTY_NAME="Ubuntu 16.04.2 LTS"
VERSION_ID="16.04"
HOME_URL="http://www.ubuntu.com/"
SUPPORT_URL="http://help.ubuntu.com/"
BUG_REPORT_URL="http://bugs.launchpad.net/ubuntu/"
VERSION_CODENAME=xenial
UBUNTU_CODENAME=xenial
\end{lstlisting}

It can be seen from the example above that Shifter makes it possible to execute applications on an arbitrary software environment that is different from the one of the host system.

\section{Extending Shifter for native GPU and MPI support}\label{sec:Extending-Shifter}

As mentioned in Section \ref{subsec:background-containers}, most of the advantages that container applications provide stem from their property of being hardware- and platform-agnostic by design.
However, when running on HPC infrastructure, this trait normally prevents the containerized application from detecting and leveraging any specialized hardware and software present in the system: examples are GPU accelerators (which are tied to specific driver versions) or vendor-specific MPI implementations that unlock the full performance of interconnect technologies like InfiniBand and Cray's Aries\texttrademark.
As a result, the benefits of migrating to a supercomputing platform are largely neutralized.

The main contributions of this work consist in extending the Shifter Runtime so that it offers user-transparent support for GPU and MPI resources available on the host system, allowing accelerated or distributed applications to achieve native performance when executed from a container.

Our extensions are implemented as part of the preparation of the software environment stage, when the Shifter Runtime sets up the container with the resources required to access the host-specific features. In the following subsections we will discuss in more detail how the support for GPU and MPI is actually implemented.

\subsection{GPU support}

As we mentioned in Section \ref{subsec:backgroud-cuda}, the method used for supporting GPU acceleration consists in bringing into a container the necessary device files, libraries, binaries, and environment variables that are needed for the correct operation of CUDA-enabled container. Furthermore, we have implemented this feature in such a way that requires little end-user intervention while only an initial system configuration is required by Shifter.

\subsubsection{Implementation details}

From a configuration point of view, there are two prerequisites for GPU support in Shifter: the host system needs to have CUDA-enabled GPUs, and the nvidia-uvm GPU driver has to be loaded prior to Shifter's execution.

At run time, the condition that triggers the activation of the native GPU support is the presence of the \texttt{CUDA\_VISIBLE\_DEVICES} variable with a valid comma-separated list of positive integers or device unique identifiers.

This choice has several reasons: first, the environment variable \texttt{CUDA\_VISIBLE\_DEVICES} is available on all CUDA ecosystems \cite{CUDACPROGRAMMING}, providing a simple and non-invasive way to check the availability of a CUDA installation without user input.
Second, the natural purpose of \texttt{CUDA\_VISIBLE\_DEVICES} is to set visibility and ordering of GPU devices for CUDA applications.
Since the host environment variables are copied into the container, the devices visible to containerized applications will be the same as for native applications.
Third, this choice respects Shifter requirement of integrating with workload managers, as some of them set the value of \texttt{CUDA\_VISIBLE\_DEVICES} upon allocating jobs, providing fine-grained control over the resources made available inside compute nodes.
If, for any reason (e.g. user request or configuration by a system administrator), the workload manager does not set \texttt{CUDA\_VISIBLE\_DEVICES} or assigns it an invalid value, Shifter does not trigger its GPU support procedure.
As an example, the SLURM workload manager can control the value of \texttt{CUDA\_VISIBLE\_DEVICES} through its Generic Resource (GRES) scheduling plugin.

When GPU support is activated, Shifter performs four operations to enable a container to execute CUDA applications:

\begin{itemize}
\item Verify that the \texttt{CUDA\_VISIBLE\_DEVICES} environment variable is present and has a valid value.

\item Add the GPU device files to the container.

\item Bind mount the CUDA driver libraries into the container.
Namely: cuda, nvidia-compiler, nvidia-ptxjitcompiler, nvidia-encode, nvidia-ml, nvidia-fatbinaryloader, nvidia-opencl.

\item Bind mount NVIDIA binaries into the container. At this stage only the command line utility NVIDIA System Management Interface (nvidia-smi) is brought into the container.
\end{itemize} 

\subsubsection{Example usage}

Consider the following example, where the CUDA SDK application \texttt{deviceQuery} is executed from a container image named \texttt{cuda-image}:

\begin{lstlisting}
    export CUDA_VISIBLE_DEVICES=0,2
    shifter --image=cuda-image ./deviceQuery
\end{lstlisting}

The application will print to screen the properties for the first and third GPU devices present in the system.

When using the SLURM workload manager to launch jobs, \texttt{CUDA\_VISIBLE\_DEVICES} is always exported into the compute node(s) environment.
GPU devices are made visible into the container by using the Generic Resource scheduling and its \texttt{--gres} command line option as follows:

\begin{lstlisting}
    srun --gres=gpu:<NumGPUsPerNode> \
    shifter --image=cuda-image ./deviceQuery
\end{lstlisting}

Therefore, by using \texttt{--gres=gpu:<NumGPUsPerNode>} Shifter transparently supports GPUs selected by the SLURM GRES plugin.

\subsubsection{On the GPU device numbering}

The GPU devices exposed inside the container are verified to be addressable by CUDA applications with device numbering starting from $0$, regardless of the actual GPU device IDs present in \texttt{CUDA\_VISIBLE\_DEVICES}. For example, if \texttt{CUDA\_VISIBLE\_DEVICES=2}, a containerized application can access the GPU using identifier $0$ with any CUDA Runtime function related to device management (e.g. \texttt{cudaSetDevice(0)}).

In other words, this makes possible for containerized applications to transparently use containers that require a single GPU even if deployed on multi-GPU systems.

\subsection{MPI support}

As mentioned in Section \ref{subsubsec:background-interfaces-mpi}, Shifter MPI support is based on the ABI-compatibility among MPI implementations that comply with the MPICH ABI compatibility initiative. The list of MPI implementations that officially adhere to this initiative is the following:

\begin{itemize}
\item MPICH v3.1 (released February 2014)
\item IBM MPI v2.1 (released December 2014)
\item Intel MPI Library v5.0 (released June 2014)
\item CRAY MPT v7.0.0 (released June 2014)
\item MVAPICH2 v2.0 (released June 2014)
\end{itemize}

In practice, the MPI libraries that adhere to the ABI-compatibility initiative fulfill the following requirements: use a specified libtool ABI string; use the same names for libraries libmpi, libmpicxx, and libmpifort; functions not covered by the MPI standard and F08 bindings are not part of the ABI; only libmpi, libmpicxx, and libmpifort libraries are valid dependencies to be used by all wrapper compilers.
Therefore, by strictly following these conventions, applications that have been built with any ABI-compatible library can later run correctly with any of the other ABI initiative library. 

\subsubsection{Implementation details}

In order to maintain the portability provided by containers while supporting site-optimized libraries, Shifter MPI support relies on the MPI implementations that provide ABI compatibility.
In practice, the MPI library that is used by a container image to allow the application to run on a commodity platform is swapped by Shifter Runtime and replaced by the ABI-compatible equivalent available on the host system.
The advantage is that the native library has been optimized for the infrastructure of the HPC system which is able to take advantage of hardware acceleration, therefore allowing the container to achieve native message-passing performance.

Shifter MPI support uses parameters that are set by the system administrator on the Runtime configuration file which specify:

\begin{itemize}

\item The full path of the host's MPI frontend shared libraries (libmpi, libmpicxx, and libmpifort);

\item The full paths to the host's shared libraries upon which the host MPI libraries depend;

\item The full paths to any configuration files and folders used by the host's MPI libraries.

\end{itemize}

Using this information, Shifter bind mounts the required host's MPI libraries and dependencies into the container environment during the setup stage, prior to application execution.
Moreover, Shifter also checks that the MPI library to be replaced is compatible with the host's MPI library: this is done by comparing the libtool ABI string of both libraries.

\subsubsection{Example usage}

Native MPI support is activated by supplying the \texttt{--mpi} command line option to the \texttt{shifter} executable.

In the following example, we use SLURM \texttt{salloc} and \texttt{srun} commands to launch a sample container that runs the \texttt{osu\_latency} application, part of the OSU Micro-Benchmarks suite~\cite{OSU-Microbenchmarks}:

\begin{lstlisting}
    salloc -N 2
    srun -n 2 --mpi=pmi2 \
      shifter --mpi --image=mpich-image \
      osu_latency
\end{lstlisting}

\section{Performance portability with Shifter}\label{sec:performanceEvaluation}

Docker containers enable users to rapidly develop and deploy applications that are self sufficient.
Nevertheless, the capability to achieve high performance is also central to the practicality of containers on the context of HPC applications.
In this section we discuss the portability and performance aspects of Shifter by presenting data for a variety of containerized scientific applications. The benchmarks have been run on a range of compute systems and software environments, completing several repetitions in order to produce statistically relevant results.

\subsection{Systems evaluated}

In order to showcase the portability and performance that can be achieved from containerized applications, we evaluate three systems that have different software environments (Linux distribution, kernel version, CUDA version, MPI implementation) and hardware configurations (CPU models, GPU models, multi-GPU setups, parallel filesystem, network interconnect).

\begin{description}[leftmargin=!]

\item [Workstation Laptop] is a Lenovo\textregistered~W540 mobile workstation equipped with an Intel\textregistered Core\texttrademark i7-4700MQ processor, 8 GB of RAM, an Nvidia\textregistered Quadro\texttrademark K110M GPU. Software versions for Linux OS, CUDA, and MPI libraries are: standard CentOS 7 (Linux kernel 3.10.0) install with CUDA 8.0, and MPICH 3.2.

\item [Linux Cluster] featuring two non-homogeneous nodes with a multi-GPU configuration: the first node contains an Intel\textregistered~Xeon\textregistered~E5-1650v3, 64 GB of RAM, one Nvidia\textregistered~Tesla\textregistered~K40m with 12 GB of on-board memory and one Nvidia\textregistered~Tesla\textregistered~K80 with 24 GB of on-board memory; the second node has an Intel\textregistered~Xeon\textregistered~E5-2650v4, 64 GB of RAM, one Nvidia\textregistered~Tesla\textregistered~K40m with 12 GB of on-board memory, and one Nvidia\textregistered~Tesla\textregistered~K80 with 24 GB of on-board memory. Both nodes are connected via an EDR Infiniband. Software versions for Linux OS, CUDA, and MPI libraries are: Scientific Linux release 7.2 (Linux kernel 3.10.0), CUDA version 7.5, and MVAPICH2 version 2.2b MPI implementation.

\item [Piz Daint] is a hybrid Cray XC50/XC40 in production at the Swiss National Supercomputing Centre (CSCS) in Lugano, Switzerland. The system compute nodes are connected by the Cray Aries interconnect under a Dragonfly topology, notably providing users access to hybrid CPU-GPU nodes.
Hybrid nodes are equipped with an Intel\textregistered~Xeon\textregistered~E5-2690v3 processor, 64 GB of RAM, a single Nvidia\textregistered~Tesla\textregistered~P100 with 16 GB  of memory. The software environment on Piz Daint is the Cray Linux Environment 6.0 UP02 (CLE 6.0) \cite{cray-cle}
using \textit{Environment Modules}~\cite{environment-modules} to provide access to compilers, tools, and applications. Software versions for Linux OS, CUDA, and MPI libraries are: CLE 6.0 (Linux kernel 3.12.60), CUDA version 8.0, and Cray's MPI Library MPT version 7.5.0.

\end{description}

\subsection{Scientific applications}

We start the study of the new capabilities introduced by our extensions to Shifter by using two scientific applications to show that:

\begin{itemize}

\item containerized applications running through Shifter can transparently access specialized resources;

\item deployed containers run without modifications on systems featuring \textit{radically} different hardware configurations and software environments.

\end{itemize}

\subsubsection{TensorFlow [CUDA-only]}

TensorFlow~\cite{abadi2016tensorflow}\cite{abadi2016tensorflow2} is a software framework providing an API to express numerical computations using data flow graphs. It also provides an implementation to run those computations on a broad array of platforms, from mobile devices to large systems with heterogeneous environments.
While usable in a variety of scientific domains, it is mainly intended for the development of machine-learning (ML) models, with a focus on deep neural networks.
The development of TensorFlow was started internally at Google Inc., and the software was released as open source in November 2015.

We use the official TensofFlow Docker image tagged \texttt{1.0.0-devel-gpu-py3} on Docker Hub.
The image is able to run GPU-accelerated TensorFlow and is built over Ubuntu 14.04, with Python 3.4.3, the NVIDIA CUDA Toolkit 8.0.44 and NVIDIA cuDNN 5.1.5 libraries.
The image also includes Google's Bazel build tool and all of TensorFlow's source code.
We arranged two test cases for evaluating the TensorFlow application:

\begin{description}[leftmargin=!,font=\normalfont\itshape]

\item [The MNIST database] \cite{mnist-dataset} is a collection of images representing handwritten digits for use in image processing and machine learning applications.
The dataset of images has been specifically prepared for the needs and purposes of machine learning research, yielding a collection of 60,000 training examples and 10,000 test examples.
Implementing a model to recognize and classify digits from the MNIST database is a classic introductory problem in machine learning.
Our test case consists in training the model implemented as a MNIST tutorial in TensorFlow's collection of community contributed models \cite{mnist-tf-tutorial}: this example is built around a simple, end-to-end, LeNet-5-like convolutional neural network.
Source code was obtained from the GitHub repository hosted at https://github.com/tensorflow/models, checking out commit \texttt{e3ad49a51e}.
We measure the wall-clock time taken by the training loop, starting immediately after parameters initialization and ending upon computation of the test error.

\item [The CIFAR-10 dataset] \cite{krizhevsky2009learning} is a collection of 60,000 small images grouped in 10 distinct classes.
The dataset is intended for training and evaluating image recognition models, counting between 50,000 training images and 10,000 test images, respectively.
Classifying the CIFAR-10 is a popular machine learning problem, and it is often featured in tutorials for ML frameworks \cite{TF-tutorial}\cite{caffe-tutorial}\cite{torch-tutorial}\cite{mocha-tutorial} or for benchmarking and comparing neural networks.
In our test case, we trained the Convolutional Neural Network (CNN) described in the TensorFlow tutorial for CNNs \cite{TF-tutorial}, which itself is a variation of the architecture by Krizhevsky \cite{krizhevsky2012imagenet} on the CIFAR-10 dataset.
The source code for the tutorial, comprising the network itself and the training harness, was obtained from the GitHub repository of community contributed TensorFlow models (https://github.com/tensorflow/models), checking out commit \texttt{e3ad49a51e}.
We run the training for 100,000 steps and report the total wall-clock time.

\end{description}

The portability of the containerized TensorFlow application was evaluated by running single-node single-GPU configuration on all three test systems: Laptop, Linux Cluster, and Piz Daint. The testing methodology used on every system: the official TensorFlow image is pulled from Docker Hub; the container was run without modification; the MNIST and CIFAR-10 workloads were executed and timed. The \texttt{nvidia-docker} program~\cite{nvidia-docker}, an extension to the Docker runtime developed by NVIDIA to provide Docker with access to the GPU, was used on the Laptop system while Shifter was used on the HPC systems.

The run times for both test cases (MNIST and CIFAR-10) are shown in Table \ref{tf-results}. We observe that by using Shifter with MPI support enabled:

\begin{enumerate}[label=\Roman*]
\item the containerized Tensorflow application is able to achieve lower run times by making use of the high-performance GPUs;
\item MNIST and CIFAR-10 test cases benefit from accessing the more powerful GPUs available on the HPC systems;
\item GPU performance portability is achieved when using the official Tensorflow container provided by a third party;
\item TensorFlow is a framework with many dependencies and elaborated build process~\cite{tf-build-from-source}, however, deploying the containerized TensorFlow is straightforward and simple;
\item the containerized application can run across HPC environments (Linux Cluster and Piz Daint) unmodified.
\end{enumerate}

\begin{table}[h]
    \centering
    \caption{Containerized TensofFlow run times (in seconds) for MNIST and CIFAR-10 tests on all three test systems.}
    \label{tf-results}
    \tabulinesep=0.9mm
    \begin{tabu}{|c|c|c|c|}
    	\hline
                 & Laptop    & Cluster  & Piz Daint \\ \hline
        MNIST    & 613       & 105     & 36        \\ \hline
        CIFAR-10 & 23359     & 8905    & 6246      \\ \hline
    \end{tabu}
\end{table}

\subsubsection{PyFR [CUDA \& MPI]}

PyFR \cite{witherdenpyfr} is a Python code for solving high-order computational fluid dynamics problems on unstructured grids.
It leverages symbolic manipulation and runtime code generation in order to run different high-performance backends on a variety of hardware platforms.
The characteristics of Flux Reconstruction make the method suitable for efficient execution on modern streaming architectures, and PyFR has been demonstrated to achieve good portability \cite{witherden2015heterogeneous} and scalability on some of the world's most powerful HPC systems: most notably, a contribution based on PyFR was selected as one of the finalists for the 2016 ACM Gordon Bell Prize \cite{vincent2016towards}.

We built a Docker image for PyFR 1.5.0 on the Laptop system, which we based on the official Docker image for Ubuntu 16.04 and added: Python 3.5.2, the NVIDIA CUDA Toolkit 8.0.44, the MPICH 3.1.4 MPI implementation, the Metis library for mesh partitioning \cite{karypis1998fast}, the specific dependencies for PyFR, and finally PyFR itself.

The correct functionality of the PyFR Docker image was validated by running it without modifications on all test systems using the 2D Euler vortex example included in the PyFR source code. \texttt{nvidia-docker} was used to run the example on the Laptop while Shifter was used on the other two HPC systems.

We selected a 3D test case representing flow over a single T106D low pressure turbine blade with a span to chord ratio of $ h/c = 1.704 $, inlet flow angle of $ 42.7{^\circ} $, outlet flow angle of $ -63.2{^\circ} $ and Reynolds number $ Re = 60,000 $.
The domain was discretized with an isotropic hexahedral mesh made of 114,265 cells and 1,154,120 points, extending for 1\% of the real blade span in the span-wise direction.
The simulation was carried out using PyFR's GPU-accelerated CUDA backend in single precision and with a fixed time-step $ dt = 9.3558\mathrm{e}{-6} [s]$ for 3,206 iterations. Due to the amount of memory required by this test case, it was not possible to run the simulation on the Laptop, so we restrict this particular test case results to the cluster systems.
For multi-node runs, each MPI rank was assigned to a single GPU.

To evaluate the portability enabled by Shifter using both GPU- and MPI-support and its influence on application scalability, we ran the test case described above on different configurations, always using the same container image.

\begin{table}[h]
    \centering
    \caption{Wall-clock times (in seconds) for PyFR on Shifter.}
    \label{table:pyfr-results}
    \tabulinesep=0.9mm
    \begin{tabu}{|c|c|c|c|c|}
        \hline
        System    & 1 GPU   & 2 GPUs   & 4 GPUs & 8 GPUs   \\ \hline
        Cluster   & 9906    & 4961     & 2509   &  -- \\ \hline
        Piz Daint & 2391    & 1223     & 620    & 322 \\ \hline
    \end{tabu}
\end{table}

The test case is setup such that for every MPI rank a single CUDA device (GPU) is assigned to it. On the Linux Cluster system runs with one, two, and four GPUs were performed while on Piz Daint the same test was run using up to 8 GPUs.
For runs on the Linux Cluster, which features two nodes and multiple GPUs per node, the devices were split among the nodes as follows:

\begin{itemize}
    \item One GPU run: a single node with one K40m GPU.
    \item Two GPUs run: two nodes with one K40m GPU each.
    \item Four GPUs run: two nodes with two GPUs each: one K40m and one K80.
\end{itemize}

All experiments were performed using Shifter with MPI- and GPU-support enabled, the best wall-clock time results obtained after $30$ repetitions are summarized in Table~\ref{table:pyfr-results}. By analyzing these results we make the following observations:

\begin{enumerate}[label=\Roman*]

\item execution times scale linearly with the increase of resources for both systems;

\item running times with single GPUs, we notice that the P100 in Piz Daint is around $4$ times faster than the K40m for this workload;

\item the four GPU run on the heterogeneous Linux Cluster setup (using both K40m and K80 GPU boards) show close to linear scaling due to a single GPU chip from the K80 board being used. Moreover, each of the the two chips on the K80 GPU board have the same architecture of a K40m GPU with similar clocks and multi-processor counts;

\item Shifter streamlines the deployment of a non-trivial Python application for diverse hardware and software environments;

\item portability is achieved as the same container is used to run across systems without modification.

\end{enumerate}

\subsection{Performance benchmarks}

We now present three benchmark applications that exclusively evaluate the performance of Shifter with: the fast network interconnect, GPU accelerators, and the parallel filesystem. At it is crucial that container portability is not at the cost of application performance. Therefore, the objective of these benchmarks is to demonstrate that:

\begin{itemize}

\item containerized applications running through Shifter can achieve the same level of performance than native applications;

\item there is a negligible overhead for accessing the specialized hardware through Shifter;

\item containerized applications running through Shifter can be deployed at scale and with a lower overhead on parallel filesystems.

\end{itemize}

\subsubsection{OSU Micro-Benchmarks [MPI-only]}

The OSU Micro-Benchmarks~\cite{OSU-Microbenchmarks} (OMB) are a widely used suite of benchmarks for measuring and evaluating the performance of MPI operations for point-to-point, multi-pair, and collective communications.
This benchmark is often used for comparing different MPI implementations and the underlying network interconnect.
In the context of this work, we use OMB to show that Shifter is able to provide the same native MPI high performance to containerized applications that use an ABI-compatible MPI implementation.

For this benchmark we used Docker and the workstation laptop to build three different containers.
The base image employed by all containers is the official CentOS 7 Linux distribution.
Each container then builds from source one of the following MPI implementation: MPICH 3.1.4 (container A), MVAPICH2 2.2 (container B), Intel MPI 2017 update 1 (container C).
Finally, for every container the OSU micro benchmark suite version 5.3.2 is built from source and then dynamically linked against each container MPI implementation.

From all the possible OSU tests availables, we have selected the OSU point-to-point latency (OSU\_latency) test.
OSU\_latency performs a ping-pong communication between a sender and a receiver where the sender sends a message and waits for the reply from the receiver.
The messages are sent repeatedly for a variety of data sizes (see column "size" of Table~\ref{table:mpi_latencies_cluster} and Table~\ref{table:mpi_latencies_pizdaint}) in order to report the average one-way latency.
Selecting this test has a dual purpose: first, it allows us to observe any possible overhead from enabling the MPI support provided by Shifter;
second, the test clearly showcases the advantages of using a hardware accelerated interconnect.

The problem setup is as follows: we run all three containers (A, B, and C) on the Linux Cluster (Infiniband interconnect) as well as Piz Daint (Aries interconnect). We then compare the results of deploying the containers through Shifter with MPI support \textit{enabled} against the same test that was natively built on each system to use the optimized MPI implementation. Finally, we rerun the containers a second time through Shifter but MPI support disabled, hence we try to use the container's MPI. 

Consider now Tables \ref{table:mpi_latencies_cluster} and \ref{table:mpi_latencies_pizdaint} that show the best results after $30$ repetitions that were obtained on the Linux Cluster and Piz Daint, respectively. It can be observed that:

\begin{enumerate}[label=\Roman*]

\item Shifter with MPI support \textit{enabled} allows all containers to achieve the same performance as the natively built test;

\item Shifter with MPI support \textit{enabled} allows all containers to transparently access the accelerated networking hardware on both the Linux Cluster and Piz Daint;

\item deploying containers using Shifter with MPI support \textit{disabled} is possible, however the containerized application does not benefit from the hardware acceleration; 

\item Shifter with MPI support \textit{disabled} leaves the responsibility to the container's MPI to use any network interface that is accessible within the container;

\item Shifter MPI-performance portability is achieved, since a container built on a workstation Laptop can run across multiple HPC systems without modification while accessing hardware accelerated interconnect.

\end{enumerate}

\begin{table}
{
\begin{center}
\tabulinesep=0.9mm
\begin{tabu} { | r | r|r|r|r|r|r|r|r |}
\hline
\multicolumn{2}{|c|}{}  &  \multicolumn{3}{c|}{Shifter MPI support} & \multicolumn{3}{c|}{Shifter MPI support} \\
\multicolumn{2}{|c|}{}  &  \multicolumn{3}{c|}{\textit{Enabled}} & \multicolumn{3}{c|}{\textit{Disabled}} \\
\hline
Size & Nat &  A &  B &  C & A &  B &  C\\
\hline
32   & 1.2   & 1.08 & 1.00 & 1.00     & 20.4 & 21.0 & 20.4\\
128  & 1.3   & 1.00 & 1.00 & 1.00     & 18.8 & 19.4 & 18.8\\
512  & 1.8   & 1.00 & 1.00 & 1.00     & 15.0 & 16.9 & 15.0\\
2K   & 2.4   & 1.00 & 1.00 & 1.00     & 29.7 & 29.9 & 29.7\\
8K   & 4.5   & 1.00 & 0.98 & 1.00     & 48.3 & 50.0 & 48.7\\
32K  & 12.1  & 1.02 & 1.02 & 1.04     & 34.5 & 34.6 & 34.5\\
128K & 56.8  & 1.00 & 1.00 & 1.01     & 26.1 & 26.4 & 23.1\\
512K & 141.5 & 0.99 & 0.99 & 1.00     & 33.3 & 33.6 & 33.5\\
2M   & 480.8 & 0.99 & 0.99 & 1.00     & 37.9 & 37.8 & 37.8\\
\hline
\end{tabu}
\captionof{table}{Results from \texttt{OSU\_latency} on the Linux Cluster:
Native runs use MVAPICH2 2.1 over Infiniband;
relative performance against native is reported for containers with (A) MPICH 3.1.4, (B) MVAPICH2 2.2, and (C) Intel MPI library using Shifter with MPI support \textit{enabled} and \textit{disabled}.}
\label{table:mpi_latencies_cluster}
\end{center}
}
\end{table}

\begin{table}
{
\begin{center}
\tabulinesep=0.9mm
\begin{tabu} { | r | r|r|r|r|r|r|r|r |}
\hline
\multicolumn{2}{|c|}{}  &  \multicolumn{3}{c|}{Shifter MPI support} & \multicolumn{3}{c|}{Shifter MPI support} \\
\multicolumn{2}{|c|}{}  &  \multicolumn{3}{c|}{\textit{Enabled}} & \multicolumn{3}{c|}{\textit{Disabled}} \\
\hline
Size & Native &  A &  B &  C & A &  B &  C\\
\hline
32   & 1.1   & 1.00 & 1.00 & 1.00     & 4.35 & 6.17 & 4.41\\
128  & 1.1   & 1.00 & 1.00 & 1.00     & 4.36 & 6.15 & 4.51\\
512  & 1.1   & 1.00 & 1.00 & 1.00     & 4.47 & 6.22 & 4.56\\
2K   & 1.6   & 1.06 & 1.00 & 1.06     & 4.66 & 5.03 & 4.04\\
8K   & 4.1   & 1.00 & 1.02 & 1.02     & 2.17 & 2.02 & 1.86\\
32K  & 6.5   & 1.03 & 1.03 & 1.03     & 2.10 & 2.17 & 1.91\\
128K & 16.4  & 1.01 & 1.01 & 1.01     & 2.63 & 2.84 & 1.95\\
512K & 56.1  & 1.00 & 1.01 & 1.01     & 2.23 & 1.78 & 1.67\\
2M   & 215.7 & 1.00 & 1.00 & 1.00     & 2.02 & 1.41 & 1.37\\
\hline
\end{tabu}
\captionof{table}{Results from \texttt{OSU\_latency} on Piz Daint:
Native runs use Cray MPT 7.5.0 over Cray Aries interconnect;
relative performance against native is reported for containers with (A) MPICH 3.1.4, (B) MVAPICH2 2.2, and (C) Intel MPI library using Shifter with MPI support \textit{enabled} and \textit{disabled}.}
\label{table:mpi_latencies_pizdaint}
\end{center}
}
\end{table}

\subsubsection{N-body benchmark [CUDA-only]}

A fast n-body simulation is included as part of the CUDA Software Development Kit~\cite{CUDA-samples}.
The n-body application simulates the gravitational interaction and motion of a group of bodies.
The application is completely implemented in CUDA and makes efficient use of multiple GPUs to calculate all-pairs gravitational interactions.
More details of the implementation can be found in~\cite{nyland2007fast}.

For this benchmark we use the official Docker image provided by NVIDIA.
It is important to note that the n-body application is already available as part of the container image, hence the containerized application can be deployed without any modification using Shifter with GPU support.
As a test case, we use the n-body benchmark to simulate $n=200,000$ bodies using double-precision floating-point arithmetic. The benchmark was run on four setups:
Laptop using the Nvidia K110M GPU;
Single-node of the Linux Cluster using a single Tesla K40m GPU;
Single-node of the Linux Cluster using a dual GPU setup with one Tesla K40m and one Tesla K80;
Single-node from Piz Daint using a single Tesla P100 GPU.
Finally, for all setups we report the gigaflops per second performance attained by native runs and container runs deployed using Shifter with GPU support enabled. (See table~\ref{table:nbody}).

The best results after $30$ repetitions show:

\begin{enumerate}[label=\Roman*]

\item containers deployed with Shifter and GPU support enabled can achieve the same performance than the natively built CUDA application;

\item the containerized application using Shifter and GPU support enabled can take full advantage of different GPU models and multiple GPUs per node;

\item GPU performance portability is achieved as the official CUDA container provided by NVIDIA is used to run the containerized n-body application across systems (Laptop, Cluster, and Piz Daint) without modification.

\end{enumerate}

\begin{table}
{
\begin{center}
\tabulinesep=0.9mm
\begin{tabu}{| c | c | c | c | c | }
    \hline
    & Laptop & Cluster & Cluster & Piz Daint \\
    & K110M & K40m & K40m \& K80 & P100 \\
    \hline
    Native & 18.34 & 858.09 & 1895.32 & 2733.01 \\
    \hline
    Container & 18.34 & 861.48 & 1897.17 & 2733.42 \\
    \hline
\end{tabu}
\caption{Giga-floating point operations per second obtained by CUDA's SDK n-body simulation application. Results report performance for native execution and running the containerized application using Shifter with GPU support enabled.}
\label{table:nbody}
\end{center}
}
\end{table}

\subsubsection{Pynamic [Parallel file system access]}

Python-based scientific applications often require a considerably long start-up time on distributed computing systems~\cite{lee2007pynamic}.
This issue is primarily due to the fact that Python performs extensive dynamic linking and loading (DLL) and parallel filesystems such as Lustre \cite{lustre} struggle to keep up with these types of loads~\cite{Jacobsen_Canon_2016}.

Pynamic~\cite{pynamic} is a benchmark designed to simulate the DLL behavior of Python-based scientific applications and measure how well an HPC system bears such types of loads.
In a nutshell, Pynamic is Python-based MPI program that performs a number of DLL operations that can be specified at build time.
This is achieved by using the Python-C API to create shared libraries that can be imported ad Python modules.

We built a Docker image for Pynamic on the Laptop system. We started with the official Docker image for Python2.7-slim (Debian Jessie) and installed MPICH 3.1.4. Finally, we built Pynamic 1.3 with the following parameters:
$495$ shared object files;
$1850$ average functions per shared object file;
Cross-module calls is enabled;
$215$ math library-like utility files;
$1850$ average math library-like utility functions;
and, $100$ additional characters on function names.

\begin{figure}[h]
	\centering
	\includegraphics[width=0.5\textwidth]{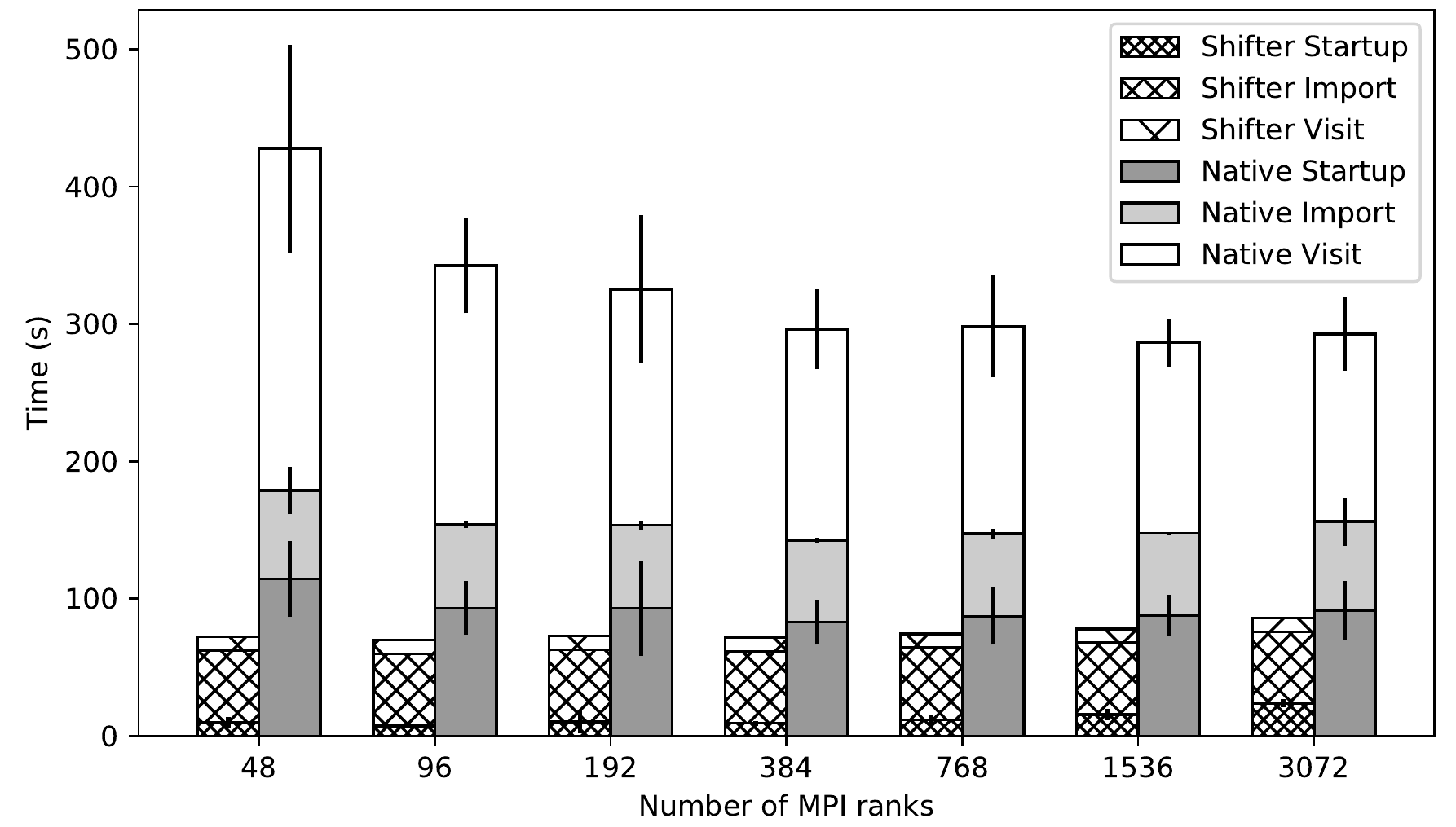}
	\caption{Performance of Pynamic running natively and on a container using Shifter on Piz Daint. Reported information shows Pynamic startup time, the time to import the Pynamic test modules, and the time to visit all functions from the imported modules. Bars report the average runtime obtained from $30$ runs while its standard deviation is shown as error bars in black.}
	\label{fig:pynamic-results}
\end{figure}

As a benchmark, we measure the wall-clock time from running the Pynamic test case for a range of MPI job sizes ($48$, $96$, $192$, $384$, $768$, $1536$, $3072$). 
Each job was executed $30$ times both natively and deployed from the container with Shifter.
The results shown in Fig.~\ref{fig:pynamic-results} present the start-up, import, and visit times of Pynamic.

From the results we observe:
\begin{itemize}

\item the reason behind the slower performance is due to the way a typical parallel filesystem works. For each DLL operation the compute node needs to request the location of the shared object to the Lustre Metadata server (MDS) and then fetch the memory block with the shared object from the Lustre Object Storage Target (OST). The main cause of the long start-up time are the repeated accesses to the MDS.

\item Shifter offers a quicker start-up time because each compute node loop mounts the container image locally. In order to access the image, the compute node needs to send only a single request to the MDS to discover the location of the image in the parallel file system.
Afterwards, the compute node only needs to fetch memory blocks directly from the OST.

\item a containerized Python  application featuring more than $3,000$ processes was deployed using Shifter showing a significant lower overhead than its native counterpart on the same parallel filesystem.
\end{itemize}

\section{Conclusions}\label{sec:conclusion}

We have discussed the challenges that users face when porting high-performance applications between HPC systems. Moreover, we have presented a solution that avoids the most common difficulties that end-users face when building and deploying scientific software on high-end computing systems while achieving high performance.
The workflow that we have presented is based on Docker and Shifter.
On the one hand, Docker images provide a self-contained and consistent software environment that addresses the portability aspects related to high-performance scientific code.
On the other hand, our work on extending Shifter allows performance portability of containers across HPC resources.

Our work provides a transparent and straightforward approach to switching dynamically linked libraries, allowing containers to load specialized versions of the ABI-compatible libraries and to access the GPUs and fast network interconnects. 

We have shown performance data on multiple benchmarks that have been run on a variety of HPC systems that support the idea of portable and high-performance containers.
In addition to the highlighted advantages, a significant gain in parallel file system access performance, which is key for large HPC deployments of Python applications like PyFR, was shown.
Moreover, the proposed container-based workflow lowers the learning curve, drastically improving user productivity and the usability of HPC systems without incurring on the loss of application performance.

\bibliographystyle{IEEEtran}
\bibliography{IEEEabrv,issue46}

% Generated by IEEEtran.bst, version: 1.14 (2015/08/26)
\begin{thebibliography}{10}
\providecommand{\url}[1]{#1}
\csname url@samestyle\endcsname
\providecommand{\newblock}{\relax}
\providecommand{\bibinfo}[2]{#2}
\providecommand{\BIBentrySTDinterwordspacing}{\spaceskip=0pt\relax}
\providecommand{\BIBentryALTinterwordstretchfactor}{4}
\providecommand{\BIBentryALTinterwordspacing}{\spaceskip=\fontdimen2\font plus
\BIBentryALTinterwordstretchfactor\fontdimen3\font minus
  \fontdimen4\font\relax}
\providecommand{\BIBforeignlanguage}[2]{{%
\expandafter\ifx\csname l@#1\endcsname\relax
\typeout{** WARNING: IEEEtran.bst: No hyphenation pattern has been}%
\typeout{** loaded for the language `#1'. Using the pattern for}%
\typeout{** the default language instead.}%
\else
\language=\csname l@#1\endcsname
\fi
#2}}
\providecommand{\BIBdecl}{\relax}
\BIBdecl

\bibitem{Ensmenger2010}
N.~L. Ensmenger, \emph{{The Computer Boys Take Over: Computers, Programmers,
  and the Politics of Technical Expertise (History of Computing)}}.\hskip 1em
  plus 0.5em minus 0.4em\relax The MIT Press, 2010.

\bibitem{mooney1990strategies}
J.~D. Mooney, ``{Strategies for supporting application portability},'' vol.~23,
  no.~11.\hskip 1em plus 0.5em minus 0.4em\relax IEEE, 1990, pp. 59--70.

\bibitem{shafer2010hadoop}
J.~Shafer, S.~Rixner, and A.~L. Cox, ``{The hadoop distributed filesystem:
  Balancing portability and performance},'' in \emph{Performance Analysis of
  Systems \& Software (ISPASS), 2010 IEEE International Symposium on}.\hskip
  1em plus 0.5em minus 0.4em\relax IEEE, 2010, pp. 122--133.

\bibitem{walters2008comparison}
J.~P. Walters, V.~Chaudhary, M.~Cha, S.~Guercio~Jr, and S.~Gallo, ``{A
  comparison of virtualization technologies for HPC},'' in \emph{Advanced
  Information Networking and Applications, 2008. AINA 2008. 22nd International
  Conference on}.\hskip 1em plus 0.5em minus 0.4em\relax IEEE, 2008, pp.
  861--868.

\bibitem{morabito2015hypervisors}
R.~Morabito, J.~Kj{\"a}llman, and M.~Komu, ``{Hypervisors vs. lightweight
  virtualization: a performance comparison},'' in \emph{Cloud Engineering
  (IC2E), 2015 IEEE International Conference on}.\hskip 1em plus 0.5em minus
  0.4em\relax IEEE, 2015, pp. 386--393.

\bibitem{Benedicic_2016}
L.~Benedicic, M.~Gila, S.~Alam, and T.~Schulthess, ``{Opportunities for
  container environments on {Cray XC30} with {GPU} devices},'' in \emph{Cray
  Users Group Conference (CUG'16)}, 2016.

\bibitem{Jacobsen_Canon_2016}
D.~M. Jacobsen and R.~S. Canon, ``{Shifter: Containers for {HPC}},'' in
  \emph{Cray Users Group Conference (CUG'16)}, 2016.

\bibitem{docker}
{Docker}, ``{\textsf{Docker}: Build, Ship, and Run Any App, Anywhere},''
  available at: \url{http://www.docker.com} (Jan. 2017).

\bibitem{lxc}
{Linux Containers project}, ``{\textsf{LXC 1.0}},'' available at:
  \url{https://linuxcontainers.org/lxc/} (Jan. 2017).

\bibitem{rkt}
{CoreOS}, ``{\textsf{rkt}: A security-minded, standards-based container
  engine},'' available at: \url{https://coreos.com/rkt/} (Jan. 2017).

\bibitem{kurtzer_2016_60736}
\BIBentryALTinterwordspacing
G.~M. Kurtzer, ``{Singularity 2.1.2 - Linux application and environment
  containers for science},'' Aug. 2016. [Online]. Available:
  \url{https://doi.org/10.5281/zenodo.60736}
\BIBentrySTDinterwordspacing

\bibitem{hale2016containers}
J.~S. Hale, L.~Li, C.~N. Richardson, and G.~N. Wells, ``{Containers for
  portable, productive and performant scientific computing},'' \emph{arXiv
  preprint arXiv:1608.07573}, 2016.

\bibitem{julian2016containers}
S.~Julian, M.~Shuey, and S.~Cook, ``{Containers in Research: Initial
  Experiences with Lightweight Infrastructure},'' in \emph{Proceedings of the
  XSEDE16 Conference on Diversity, Big Data, and Science at Scale}.\hskip 1em
  plus 0.5em minus 0.4em\relax ACM, 2016, p.~25.

\bibitem{Felter_2015}
W.~Felter, A.~Ferreira, R.~Rajamony, and J.~Rubio, ``{An updated performance
  comparison of virtual machines and linux containers},'' in \emph{Performance
  Analysis of Systems and Software (ISPASS), 2015 IEEE International Symposium
  On}.\hskip 1em plus 0.5em minus 0.4em\relax IEEE, 2015, pp. 171--172.

\bibitem{boettiger2015introduction}
C.~Boettiger, ``{An introduction to Docker for reproducible research},''
  \emph{ACM SIGOPS Operating Systems Review}, vol.~49, no.~1, pp. 71--79, 2015.

\bibitem{MPI30}
{Message Passing Interface Forum}, ``{\textsf{MPI}: A Message-Passing Interface
  Standard. Version 3.0},'' September 21 2012, available at:
  \url{http://www.mpi-forum.org} (Jan. 2017).

\bibitem{MPIABI}
{MPICH}, ``{MPICH ABI Compatibility Initiative},'' November 2013, available at:
  \url{https://www.mpich.org/abi/} (Jan. 2017).

\bibitem{CUDACPROGRAMMING}
{NVIDIA}, ``{CUDA C Programming Guide},'' September 2016, available at:
  \url{http://docs.nvidia.com/cuda} (Jan. 2017).

\bibitem{deal2016hpc}
S.~J. Deal, ``Hpc made easy: Using docker to distribute and test trilinos,''
  2016.

\bibitem{OSU-Microbenchmarks}
{Network-Based Computing Laboratory}, ``{OSU Microbenchmarks},'' available at:
  \url{http://mvapich.cse.ohio-state.edu/benchmarks/} (Jan. 2017).

\bibitem{cray-cle}
{CRAY}, ``{XC Series System Administration Guide (CLE 6.0)},'' available at:
  \url{https://pubs.cray.com} (Mar. 2017).

\bibitem{environment-modules}
{Environment Modules project}, ``{Environment Modules open source project},''
  available at: \url{http://modules.sourceforge.net/} (Mar. 2017).

\bibitem{abadi2016tensorflow}
M.~Abadi, A.~Agarwal, P.~Barham, E.~Brevdo, Z.~Chen, C.~Citro, G.~S. Corrado,
  A.~Davis, J.~Dean, M.~Devin \emph{et~al.}, ``{Tensorflow: Large-scale machine
  learning on heterogeneous distributed systems},'' \emph{arXiv preprint
  arXiv:1603.04467}, 2016.

\bibitem{abadi2016tensorflow2}
M.~Abadi, P.~Barham, J.~Chen, Z.~Chen, A.~Davis, J.~Dean, M.~Devin,
  S.~Ghemawat, G.~Irving, M.~Isard \emph{et~al.}, ``{TensorFlow: A system for
  large-scale machine learning},'' in \emph{Proceedings of the 12th USENIX
  Symposium on Operating Systems Design and Implementation (OSDI). Savannah,
  Georgia, USA}, 2016.

\bibitem{mnist-dataset}
Y.~LeCun, L.~Bottou, Y.~Bengio, and P.~Haffner, ``{Gradient-based learning
  applied to document recognition},'' \emph{Proceedings of the IEEE}, vol.~86,
  no.~11, pp. 2278--2324, 1998.

\bibitem{mnist-tf-tutorial}
{The TensorFlow Authors}, ``{Simple, end-to-end, LeNet-5-like convolutional
  MNIST model example},'' available at:
  \url{https://github.com/tensorflow/models/blob/master/tutorials/image/
  mnist/convolutional.py} (Jan. 2017).

\bibitem{krizhevsky2009learning}
A.~Krizhevsky and G.~Hinton, ``{Learning multiple layers of features from tiny
  images},'' 2009.

\bibitem{TF-tutorial}
{TensorFlow}, ``{Convolutional Neural Networks},'' available at:
  \url{https://www.tensorflow.org/tutorials/deep\_cnn} (Jan. 2017).

\bibitem{caffe-tutorial}
{Caffe project}, ``{Alex’s CIFAR-10 tutorial, Caffe style},'' available at:
  \url{http://caffe.berkeleyvision.org/gathered/examples/cifar10.html} (Jan.
  2017).

\bibitem{torch-tutorial}
{Sergey Zagoruyko}, ``{$92.45\%$ on CIFAR-10 in Torch},'' available at:
  \url{http://torch.ch/blog/2015/07/30/cifar.html} (Jan. 2017).

\bibitem{mocha-tutorial}
{Mocha project}, ``{Alex’s CIFAR-10 tutorial in Mocha},'' available at:
  \url{http://mochajl.readthedocs.io/en/latest/tutorial/cifar10.html} (Jan.
  2017).

\bibitem{krizhevsky2012imagenet}
A.~Krizhevsky, I.~Sutskever, and G.~E. Hinton, ``{Imagenet classification with
  deep convolutional neural networks},'' in \emph{Advances in neural
  information processing systems}, 2012, pp. 1097--1105.

\bibitem{nvidia-docker}
{NVIDIA}, ``{Build and run Docker containers leveraging NVIDIA GPUs},''
  available at: \url{https://github.com/NVIDIA/nvidia-docker} (Feb. 2017).

\bibitem{tf-build-from-source}
{TensorFlow}, ``{Installing TensorFlow from Sources},'' available at:
  \url{https://www.tensorflow.org/install/install\_sources} (Jan. 2017).

\bibitem{witherdenpyfr}
F.~Witherden, A.~Farrington, and P.~Vincent, ``{PyFR: An Open Source Framework
  for Solving Advection-Diffusion Type Problems on Streaming Architectures}.''

\bibitem{witherden2015heterogeneous}
F.~D. Witherden, B.~C. Vermeire, and P.~E. Vincent, ``{Heterogeneous computing
  on mixed unstructured grids with PyFR},'' \emph{Computers \& Fluids}, vol.
  120, pp. 173--186, 2015.

\bibitem{vincent2016towards}
P.~Vincent, F.~Witherden, B.~Vermeire, J.~S. Park, and A.~Iyer, ``{Towards
  green aviation with python at petascale},'' in \emph{Proceedings of the
  International Conference for High Performance Computing, Networking, Storage
  and Analysis}.\hskip 1em plus 0.5em minus 0.4em\relax IEEE Press, 2016, p.~1.

\bibitem{karypis1998fast}
G.~Karypis and V.~Kumar, ``{A fast and high quality multilevel scheme for
  partitioning irregular graphs},'' \emph{SIAM Journal on scientific
  Computing}, vol.~20, no.~1, pp. 359--392, 1998.

\bibitem{CUDA-samples}
{NVIDIA}, ``{CUDA SAMPLES Reference Manual. January 2017},'' available at:
  \url{http://docs.nvidia.com/cuda/cuda-samples/} (Feb. 2017).

\bibitem{nyland2007fast}
L.~Nyland, M.~Harris, J.~Prins \emph{et~al.}, ``{Fast N-Body Simulation with
  CUDA},'' \emph{GPU gems}, vol.~3, no.~31, pp. 677--695, 2007.

\bibitem{lee2007pynamic}
G.~L. Lee, D.~H. Ahn, B.~R. de~Supinski, J.~Gyllenhaal, and P.~Miller,
  ``{Pynamic: the Python dynamic benchmark},'' in \emph{Workload
  Characterization, 2007. IISWC 2007. IEEE 10th International Symposium
  on}.\hskip 1em plus 0.5em minus 0.4em\relax IEEE, 2007, pp. 101--106.

\bibitem{lustre}
{Lustre project}, ``{Lustre filesystem},'' available at:
  \url{http://lustre.org} (Mar. 2017).

\bibitem{pynamic}
{Gregory Lee, Dong Ahn, John Gyllenhaal, Bronis de Supinski}, ``{Pynamic: The
  Python Dynamic Benchmark},'' available at:
  \url{https://codesign.llnl.gov/pynamic} (Mar. 2017).

\end{thebibliography}

\end{document}